\documentstyle[12pt]{article}
 \input{epsf}
\newcommand{\N}{{\rm I \hspace{-0.52ex} N}}
\newcommand{\unop}{{\rm 1 \hspace{-0.52ex} I}}
\begin{document}
\setlength\textheight{8.75in}
\newcommand{\be}{\begin{equation}}
\newcommand{\ee}{\end{equation}}
\title{PT-Symmetric, Quasi-Exactly Solvable matrix Hamiltonians}
\author{Yves Brihaye,\\
{\it
 Universit\'e de Mons-Hainaut, Fac. Sciences
B-7000 Mons, Belgium},\\
Ancilla Nininahazwe, \\
{\it Universit\'e du Burundi, Fac. Sciences, B.P.2700,Bujumbura},\\
and\\
Bhabani Prasad Mandal,\\
{\it Department of Physics, Banaras Hindu University, Varanasi-221005,India.}}
 \date{\today}
 \maketitle
 \thispagestyle{empty}
 \begin{abstract}
Matrix quasi exactly solvable operators are considered and new
conditions are determined to test whether a matrix differential operator
possesses a or several finite dimensional invariant vector spaces.
New examples
of $2\times 2$-matrix quasi exactly solvable operators are constructed
with the emphasis set on PT-symmetric Hamiltonians.
 \end{abstract}
 \medskip
 \medskip
 \newpage
 
\section{Introduction}
Quasi exactly solvable (QES) operators refer to a class of linear operators
(typically of Schr\"odinger type) which preserve a finite dimensional
subspace of the Hilbert space on which they act \cite{tur1,tur2,ts}.
Is most of the examples
known, QES operators can be transformed into operators preserving a space
of polynomials of given degree after a suitable change of variable and
change of functions (also called "gauge transformations"). The operators
preserving a space of polynomials therefore play an important role in
the study of QES operators.
In the case of scalar equations, the gauge transformation consists
in factorizing the ground state out of the wave function,
the change of variable can be performed in a straighforward way
(at least formally because it leads usually to elliptic
integrals).         

In the case of coupled equations, where the operators appear
in the form of a matrix whose components are differential
operators,
the construction of the gauge transformation setting the operator
in a form which manifestly preserves a vector space where component
are polynomials, appear more tricky, see e.g. \cite{kb,bh}.

In the second section of this paper, we establish a set of algebraic
conditions to test wether $n \times n$ matrix-valued operators
of a certain type
preserve a vector space of $n$-uple of polynomials with component of definite
degrees. We work with $2\times 2$ matrices but the method can be extended
to higher dimensions.     This new method was tested  on all 
QES known matrix equations.
In the other sections, we take advantage of these conditions and construct
several new families of QES systems where the emphasis is set on
PT-symmetric invariance. This original issue for the mathematical
framework of quantum mechanics was proposed in \cite{bender} and
developped is several subsequent papers but, to our knowledge, it has not
been studied in the context of coupled systems of Schrodinger equations.

In Sect. 3 we propose
several  matrix extentions
of the Razhavi operator.
Scalar Razhavi-types of potentials
were considered recently to produce examples of PT-invariant,
non-hermitian potentials with real eigenvalues \cite{khare,khare1}.
Here, we develop matrix extensions of them both with  trigonometric
and hyperbolic potentials.

In Sect. 4 we obtain a matrix generalisation
of the  QES example of PT-symmetric hamiltonian with an anharmonic
potential of degree four \cite{bender} and reconsidered recently
\cite{zn}. Finally, in Sect. 5, we show how the problem
of section 4 can be transformed into
a system of recurence equations in the spirit of \cite{bd}.

\section{Matrix QES Operator}
In this section, we propose a general test to check
whether a $2\times 2$ matrix
differential operator  $H$ (depending of the variable $x$ and of the
derivatives $d^n/dx^n$),
preserves a vector space whose components are polynomials of suitable degrees
in $x$.
We will consider a family of operators $H$ which can be decomposed
according to
\be
\label{generalform}
H = H_1 + H_0 + \dots,
\ee
The diagonal components of  $H_1$
are differential operators of degree $1$
(i.e acting on a generic polynomial of degree $n$ in $x$
they increase the degree by one unit) and  the off-diagonal elements
$(H_1)_{12}$ and $(H_1)_{21}$ are respectively
proportional to $x^\delta$ and $x^{\delta^\prime}$,
with $\delta=0,1,2$ and $\delta^\prime\equiv 2-\delta$.
The operators $H_0, \cdots$ have lower degrees in all their components
than the corresponding components in $H_1$. The dots in (\ref{generalform})
represent operators with lower degree ordered according to the same rule.

Most of the QES matrix operators known \cite{bh, brihaye,zhdanov} can be
reduced to the form above after a suitable gauge transformation
and change of variable.
The different components of $H_1$, $H_0$ contain several
constant parameters which label the physical coupling constants
in case where $H$ is an Hamiltonian operator.

Now, we try to obtain the conditions of $H_1$ and $H_0$ (i.e. conditions
on the coupling constants) such that the operator $H$ possesses
a finite dimensional invariant subspace of polynomials of the form
\be
{\cal  V} = {\rm span} \Biggl\{\left(\matrix{p_n\cr
                                           q_m }\right)
                      \Biggr\}  \ \
                 , \ \  n \in \N \ \ , \ \ m = n-\delta+1
\ee
where $p_n, q_m$ denote polynomials of degree $n, m$ in the variable $x$.
For a generic vector in $\cal V$ of the form
\be
 \psi = \left(\matrix{\alpha_0x^n\cr
                 \beta_0x^{n-\delta+1}\cr}\right) 
                 + \left(\matrix{\alpha_1x^{n-1}\cr
                 \beta_1x^{n-\delta}\cr}\right) + \dots,
\ee
where $\alpha_i,\beta_i$ are complex parameters,
the vector $H \psi$ can be decomposed according to
\begin{eqnarray}
&H \psi&={\rm diag}(x^{n+1},x^{n-\delta+2}) M_1\left(\matrix{\alpha_0\cr\beta_0\cr}\right)\nonumber\\
&                &+\Biggl({\rm diag}(x^{n},x^{n-\delta+1})\tilde {M_1}\left(\matrix{\alpha_1\cr
                 \beta_1\cr}\right) + {\rm diag}(x^{n},x^{n-\delta+1})M_0\left(\matrix{\alpha_0\cr
                 \beta_0\cr}\right)\Biggr)\nonumber\\
&                &+ {\rm terms \ of \ lower \ degrees} \  \ ,
\end{eqnarray}
where  the constant $2\times 2$
 matrices $M_1, \tilde {M_1}$ and $M_0$ can be obtained after a simple
 algebra.

The necessary conditions for $\cal V$ to contain
 an invariant vector space of the operator $H$ read
\begin{eqnarray}
&(i) \ \ &M_1 \left(\matrix{\alpha_0\cr\beta_0\cr}\right)
= \left(\matrix{0\cr 0\cr}\right)    \\
&(ii) \ \  &\tilde {M_1}\left(\matrix{\alpha_1\cr
                 \beta_1\cr}\right) + M_0\left(\matrix{\alpha_0\cr
                 \beta_0\cr}\right) \propto
                 \left(\matrix{\alpha_0\cr\beta_0\cr}\right)
                 \end{eqnarray}
where the second condition has to be fullfilled
irrespectively of the values $\alpha_1, \beta_1$.

The condition $(i)$ implies ${\rm det} M_1 = 0$ and the vector
$(\alpha_0, \beta_0)^t$ to be a
 zero-eigenvalue eigenvector of $M_1$.
This fixes the  relative coefficient
of the terms of highest degree in ${\cal V}$ (see  Eq.(2)).
The condition $(ii)$ can  be fullfilled only if
the following  conditions hold
\be  
\label{mat}
 (ii') \ \ \ {M_0}\left(\matrix{\alpha_0\cr
                 \beta_0\cr}\right) = \Lambda \left(\matrix{\alpha_0\cr
                 \beta_0\cr}\right) \ \ , \ \
 {\tilde M_1^t}\left(\matrix{-\beta_0\cr
                 \alpha_0\cr}\right) =  \left(\matrix{0\cr
                 0\cr}\right)   \ \ ,
\ee
where $M^t$ means the transpose matrix of $M$.


The conditions (i),(ii') allow to reconstruct in a systematic
way the invariant vector spaces of all QES operators
presented e.g. in \cite{zhdanov,bh,brihaye}; in particular the conditions
on the different parameters and the relevant changes of variable
now emerge in terms of elementary algebra on matrices.

In order to illustrate this method, we reconstruct
the invariant vector space of the QES Hamiltonian
\cite{zhdanov,brihaye}
\be
H(y) = -\frac{d^2}{dy^2}{\unop}_2 + M_6(y),
\ee
where $M_6(y)$ is a $2\times 2$ hermitian matrix of the form
\be
M_6(y)=\{4p_2^2y^6 + 8p_1p_2y^4 + (4p_1^2 - 8mp_2 + 2(1 - 2\epsilon)p_2)y^2\}{\unop}_2  + (8p_2y^2 + 4p_1)\sigma_3-8mp_2\kappa_0\sigma_1
\ee
It is known that after the usual "gauge transformation" of $H(y)$ with a factor
\be
\phi(y) = y^\epsilon\exp-\{\frac{p_2}{2}y^4 + p_1y^2\}  \ \ , \ \ \epsilon=0,1
\ee
and the change of variable $x=y^2$, the new operator $\tilde H(x)$ is obtained
\be
\tilde H(x)= \phi^{-1}(y)H(y)\phi(y)\vert_{y=\sqrt x}
\ee
For simplicity we assume $\epsilon=0, p_1=0$  in the following.
The operator obtained $\tilde H(x)$ then reads
\be
\tilde H(x) = (-4x\frac{d^2}{dx^2}-2\frac{d}{dx}){\unop}_2 + 8p_2\left(\matrix{J_+(m-2)&0\cr
                0 &J_+(m)\cr}\right)-8mp_2\kappa_0\sigma_1
\ee
with $J_+(m) \equiv x^2 d_x - m x$.
It can be decomposed along the lines of Eq.(\ref{generalform})~:
\be
\tilde H(x) = H_1 + H_0 + H_{-1},
\ee
with
\begin{eqnarray}
&H_1 &= 8 p_2 \left(\matrix{J_+(m-2)&- m \kappa_0\cr
                0 &J_+(m)\cr}\right),\nonumber\\
&H_0 &= 0,\nonumber\\
& H_{-1}&= (-4x\frac{d^2}{dx^2}-2\frac{d}{dx}){\unop}_2 - 8mp_2\kappa_0\sigma_-              
\end{eqnarray}
In this case,  $(H_1)_{12}$ is a constant(i.e. $\delta=0$),
while $(H_1)_{21}=0$, in addition the operatot $H_0$ is zero.
The invariant vector space to be looked for is of the form
\be
 \psi = \left(\matrix{\alpha_0x^{m-1}+\alpha_1x^{m-2}+\dots\cr
                 \beta_0x^m + \beta_1x^{m-1}+\dots\cr}\right)
\ee
The determinant of the matrix $M_1$ is trivially zero
and the condition $(i)$ implies $\frac{\alpha_0}{\beta_0}=m\kappa_0$.
The first conditions $(ii')$ is trivial since $M_0=0$
(as a consequence of $H_0=0$). Finally, the second condition 
$(ii')$  can be easily checked~:
\begin{eqnarray}
&\tilde {M_1}\left(\matrix{\alpha_1\cr
                \beta_1\cr}\right)
                &= -8p_2\beta_1\left(\matrix{m\kappa_0\cr
                1\cr}\right)
                 = -8p_2\beta_1\left(\matrix{\frac{\alpha_0}{\beta_0}\cr
                1\cr}\right)\nonumber\\
&-8p_2\beta_1\beta_0\left(\matrix{m\kappa_0\cr
                1\cr}\right)&=-\beta_1\left(\matrix{\alpha_0\cr
                \beta_0\cr}\right)
\end{eqnarray}
In the following section, we will present
several examples of QES matrix operator based on extensions
of the scalar Razavi potential.

\section{\bf PT invariant non-hermitian matrix Hamiltonian }
In \cite{khare,khare1} PT-invariant models based
on the scalar Razhavi potential are analyzed with the emphasis set
on the reality properties of the spectrum. This can be done partly in an
analytical way because the potentials considered are QES.
The authors considerd both, hyperbolic and trigonometric cases, invoking
an anti-isospectral transformation \cite{ush} to relate the
spectra of both types.
Here we will consider matrix extensions of these equations and see that
several form of the non diagonal elements $H_{12}$ and $H_{21}$ can
lead to QES operators. We will first consider periodic potentials, formulated
in terms of trigonometric functions. Then an example involving 
elliptic functions for the potentials will be presented.

\subsection{Trigonometric case}
From the unidimensional potential studied in \cite{khare},
we will build a family
PT invariant matrix Hamiltonian and use the technique developped
the previous section to check its quasi exactly solvability.
We start from a general  Hamiltonian of the  form

\be
\label{trigono}
H = \left(\matrix{-\frac{d^2}{dx^2}+(\rho\cos 2x-iM)^2 + A & H_{12}\cr
H_{21}&-\frac{d^2}{dx^2}+(\rho\cos 2x-i\tilde M)^2 + \tilde A \cr}\right),
\ee
where $\rho$ is a free real parameter and $A, \tilde A, M, \tilde M$ 
are constant to be specified. There are several forms of $H_{12}$, $H_{21}$
which lead to QES operators. 
One can assume  $\tilde M>M$ without loosing generality.
The general properties of the diagonal
component of $H$ and of trigonometric fonctions will reveal that
QES operators can be constructed by choosing $H_{12}$ according to
one of the following form
\be
   H_{12} = C \cos 2 x + D \ \ {\rm or} \ \
   H_{12} = C \cos  x  \ \ {\rm or}   \ \
   H_{12} = C \sin  x  \ \ {\rm or}  \ \
   H_{12} = C \cos  x \sin x
   \label{diagonalterm}
\ee
and similar forms respectively for $H_{21}$ with, however,
a priori independent coupling constants for $C$ and $D$.

In order to reveal the algebraic properties of this family of
operators, it is convenient to perform a first
gauge transformation according to
\begin{eqnarray}
&\tilde H &= e^{-\theta\cos 2x}
\left(\matrix{ z^{-\epsilon} (1-z)^{-\phi}&0 \cr
                0 & z^{- \tilde \epsilon} (1-z)^{-\tilde \phi}\cr}\right)
H
e^{\theta\cos 2x}
\left(\matrix{ z^{\epsilon} (1-z)^{\phi}&0 \cr
                0 & z^{\tilde \epsilon} (1-z)^{\tilde \phi}\cr}\right)
,\nonumber\\
&         &=\left(\matrix{\tilde H_{11}&\tilde H_{12}\cr
                \tilde H_{21} & \tilde H_{22}\cr}\right),
\end{eqnarray}

where $z= (\cos 2x+1)/2$. Further choosing the parameter $\theta$,
according to $ \theta= i\frac{\rho}{2}$ the components of $\tilde H$
are obtained after an algebra:
\begin{eqnarray}
&\tilde H_{11}&= -4z(1-z)\frac{d^2}{dz^2} +
             2(2z-1 -4(1-z)\epsilon + 4 \phi z)\frac{d}{dz} 
             + \rho^2 - M^2 + 8 \phi\epsilon + 2 \epsilon + 2 \phi + A\nonumber \\
    & &- 8i \rho (z(1-z) \frac{d}{dz} + \epsilon(1-z)-\phi z + \frac{M-1}{4}(2z-1))
 \nonumber\\
&\tilde H_{12}&= z^{\tilde \epsilon - \epsilon} (1-z)^{\tilde \phi - \phi} H_{12} ,\nonumber\\
&\tilde H_{21}&= z^{\epsilon - \tilde \epsilon} (1-z)^{\phi - \tilde \phi} H_{21},\nonumber\\
&\tilde H_{22}&= -4z(1-z)\frac{d^2}{dz^2} +
             2(2z-1 -4(1-z)\tilde \epsilon + 4 \tilde \phi z)\frac{d}{dz}
             + \rho^2 - M^2 + 8 \tilde\phi\tilde\epsilon + 2 \tilde\epsilon 
             + 2 \tilde \phi + \tilde A  \nonumber \\
    & &- 8i \rho (z(1-z) \frac{d}{dz} + \tilde \epsilon(1-z)-\phi z 
    + \frac{\tilde M-1}{4}(2z-1))
\end{eqnarray}
and where we have neglected the singular terms of the form
\be
      \frac{1-z}{z} 2\epsilon (2 \epsilon -1) 
      + \frac{z}{1-z} 2\phi (2 \phi -1)
\ee
in $H_{11}$ (and a similar terms with $\epsilon \to \tilde \epsilon$ ,
$\phi \to \tilde \phi$ in $H_{22}$) since we assume from now on
\be
\epsilon(2\epsilon-1) = \phi(2\phi-1)=\tilde \epsilon(2\tilde \epsilon-1)
= \tilde \phi(2 \tilde \phi-1)=0
\label{values}
\ee

The different choices for $H_{12}$ proposed in Eq.(\ref{diagonalterm})
now appear to be natural since they will automatically lead to a polynomial
expressions in $z$ when the choice of the parameters
$\epsilon, \tilde \epsilon, \phi, \tilde \phi$ is done
according to Eq. (\ref{values}).
In the following, we will analyze in details the case $H_{12}= C \sin x \cos x$.
The algebraisation corresponding to the three other cases can be done similarly.

In this case, the  possible values for the parameters
$\epsilon, \tilde \epsilon, \phi, \tilde \phi$ allow for four algebraisation,
for the wave function $\psi = (\psi_1, \psi_2)$, namely~:
\begin{eqnarray}
&{\rm type} \  i \ : \ &\psi = (p_n , \sin x \cos x  q_{n-1} ) \nonumber  \\
&{\rm type} \ ii \ : \  &\psi = ( p_{n-1} \sin x \cos x,  q_n  ) \nonumber  \\
&{\rm type} \ iii \ :\ &\psi = ( p_n \sin x,  q_{n} \cos x)  \nonumber  \\
&{\rm type} \ iv \ : \ &\psi = ( p_n \cos x,  q_{n} \sin x) \nonumber
\end{eqnarray}
where $p_n,q_n$, etc. denote polynomials of degree $n$ in the variable $z$.

 Acting on an eigenfunction of type (i),
the conditions for algebraic solutions are $\epsilon = \phi = 0$,
$\tilde \epsilon = \tilde \phi = 1/2$.
The operator  $\tilde H$ can then be decomposed according to
the prescription of Sect. 2, leading to~:
\be
\tilde H = \tilde H_1 + \tilde H_0 + \tilde H_{-1},
\ee
whith
\be
\tilde H_1 = \left(\matrix{8 i \rho (z^2\frac{d}{dz}-(\frac{M-1}{2}) z)& -Cz^2\cr
 \tilde C& 8 i \rho(z^2\frac{d}{dz}-(\frac{\tilde M -3}{2})z\cr}\right),
\ee
\be
\tilde H_0 = (4z^2\frac{d^2}{dz^2}+(4-8i\rho)z\frac{d}{dz}+\rho^2)\unop_2  +
\left (\matrix{
A'  & C z \cr
0 &
8 z \frac{d}{dz} + \tilde A'  \cr}\right)
\ee
and
\be
\tilde H_{-1} = \left ( \matrix{ -4z\frac{d^2}{dz^2} -2\frac{d}{dz} & 0 \cr
 0&-4z\frac{d^2}{dz^2} -6\frac{d}{dz} \cr} \right ).
\ee
with 
\be
A' = A-M^2 + 2i\rho(M-1)\ , \ \tilde A' = \tilde A + 4 -{\tilde M}^2+2i\rho(\tilde M-3)
\ee

Using  the parametrisation corresponding to type (i) for the wave function,
we can easily obtain the form of the matrices $M_1, \tilde M_1, M_0$
and the conditions on the parameters leading to QES operators.
In the present case, we got
\be
\label{cond1}
   M + \tilde M = 4 n \ \ , \ \ (1-4 n^2) + M \tilde M = \frac{C \tilde C}{16 \rho^2}
\ee
The condition involving $M_0$ fixes the difference between the constant
$A , \tilde A$, namely
\be
\label{cond2}
    A-\tilde A = M^2- {\tilde M}^2
\ee
Considering the parametrisation corresponding to type (ii) for the wave function,
one can easily obtain $\epsilon = \phi =1/2 $,
$\tilde \epsilon = \tilde \phi = 0$ and the associated operator $\tilde H$. The action 
of $\tilde H$ on an eigenfunction of the type (ii) gives after an algebra the matrices 
$M_1, \tilde M_1, M_0$ which lead to the same QES conditions found 
in the previous case as given by the Eq.(\ref{cond1}) and the Eq.(\ref{cond2}).

This time, the wave functions of the type (iii) and of the type (iv) correspond
respectively to $\epsilon = \tilde \phi  = 0$, $\tilde \epsilon =\phi = 1/2 $
and $\epsilon = \tilde \phi  = 1/2 $, $\tilde \epsilon =\phi = 0$. After some
algebra, one can find the corresponding operators $\tilde H$ and also the matrices
$M_1, \tilde M_1, M_0$ are deduced. For these two types (iii) and (iv), we got 
the three same QES conditions 
\be
\label{cond2}
   M + \tilde M = 4 n + 2 \ \ , \ \  M \tilde M - 4n(n+1) = \frac{C \tilde C}{16 \rho^2}
\ee
The condition involving $M_0$ fixes the difference between the constant
$A , \tilde A$, namely
\be
 A-\tilde A = M^2- {\tilde M}^2.
\ee
It is found  that this QES condition is the same for all four types
of the wave function.

\begin{figure}
\epsfysize= 9cm
\epsffile{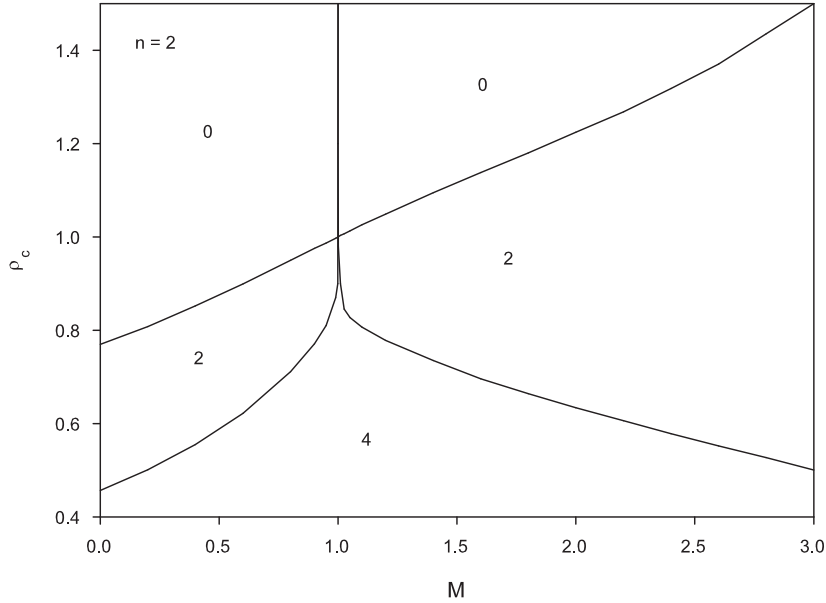}
\caption{\label{Fig.1} The critical value of $\rho$ as a function
of the coupling constant $M$ for the type $i$ solution and $n=2$.
The integers label the number of real algebraic eigenvalues}
\end{figure}

A consequence of these results is that the equations (\ref{trigono})
admits a double algebraisation. Solutions of the types $i$ and $ii$ 
exist if the
condition (\ref{cond1}) is fulfilled 
and solution of the  types $iii$ and $iv$ if the condition
(\ref{cond2}) holds.

 \subsection{Some properties of the spectrum}
 We studied the algebraic eigenvalues of the equation (\ref{trigono})
 for the solution of the type $i$ and for $n=1,2$.
 The invariant vector space has dimension $2n$ keeping into account
 that the condition (5) imposes a constraint on the polynomials.
 In the case $n=1$ the two algebraic eigenvalues have the form
 \be
       E =  \rho^2 + 2 \pm \sqrt{1 - \rho^2 (1+M )^2} \ .
 \ee
Showing that the algebraic eigenvalues are real only for $|\rho| < 1/|1+M|$.
Similar features are observed namely in \cite{khare},\cite{khare1}.
Namely the eigenvalues come out to be real or in complex degenerate pairs.
For $n=2$ the polynomial giving the four algebraic eigenvalues for $E$
is real but rather involved and the solution
cannot be expressed in a closed form for generic values
of $M, \rho$. In the case $M=1$ and $M=3$, however,
 we could find explicit solutions~:
$$
   M = 1 \ \ , \ \ E = 4 + \rho^2 \ \ (2 \ {\rm times}) \ \ , \ \
                   E = 8 + \rho^2 \pm 8 \sqrt{1-\rho^2}
$$
$$
 M = 3 \ \ , \ \ E = 10 + \rho^2  \pm \sqrt{9 - 4 \rho^2} \ \ , \ \
                   E = 2 + \rho^2 \pm 2 \sqrt{1- 4\rho^2}
$$
The plane $M, \rho$ is partitionned into regions admitting
$4$,$2$ or $0$ algebraic eigenvalues. The critical values of 
$\rho_c$ are presented as functions of the parameters $M$
for $0 \leq M \leq 3$. We note in particular that one of the
critical line become infinite in the limit $M \to 1$, indicating that
in this case there are two real eigenvalues for $\rho > 1$.

\subsection{Hyperbolic case}
The construction of previous section can also be realized
for the case where the trigonometric functions
entering in the potentials are replaced by their elliptic counterpart.
The discussion of the different algebraisations turn out to be
the same. Here however, we will study in detail the algebraic
properties of the operator given by
\be
H = \left(\matrix{-\frac{d^2}{dx^2}-(\rho\cosh 2x-iM)^2 &C(\cosh 2x-1)+\tilde C\cr
                 D(\cosh 2x-1)+\tilde D&-\frac{d^2}{dx^2}-(\rho\cosh 2x-i\tilde M)^2 \cr}\right),
\ee
where $\rho$ is a free real parameter.
One can assume  $\tilde M>M$ without loosing generality.
The gauge transformation is performed as follows
\begin{eqnarray}
&\tilde H &= \exp(-\theta\cosh 2x)H\exp(\theta\cosh 2x),\nonumber\\
&         &=\left(\matrix{\tilde H_{11}&\tilde H_{12}\cr
                \tilde H_{21} & \tilde H_{22}\cr}\right),
\end{eqnarray}
On further substituting $z=\cosh 2x-1$ and fixing the
constant $\theta$ by means of  $\theta=\frac{i\rho}{2}$,
the different components of $\tilde H$ read
\begin{eqnarray}
&\tilde H_{11}&=-4z(z+2)\frac{d^2}{dz^2}-4(z+1)\frac{d}{dz}-8i\rho z\frac{d}{dz}-\rho^2-4i\rho(z^2\frac{d}{dz}-\frac{M-1}{2}z)\nonumber\\
&  &+2i\rho(M-1)+M^2,\nonumber\\
&\tilde H_{12}&= Cz+\tilde C,\nonumber\\
&\tilde H_{21}&= Dz+\tilde D,\nonumber\\
&\tilde H_{22}&=-4z(z+2)\frac{d^2}{dz^2}-4(z+1)\frac{d}{dz}-8i\rho z\frac{d}{dz}-\rho^2-4i\rho(z^2\frac{d}{dz}-\frac{\tilde M-1}{2}z)\nonumber\\
& &+2i\rho(\tilde M-1)+{\tilde M}^2.
\end{eqnarray}

Decomposing now the operator $\tilde H$ according to Eq.(\ref{generalform}),
we obtain
\be
\tilde H_1 = \left(\matrix{-4i\rho(z^2\frac{d}{dz}-Nz)&Cz\cr
                Dz& -4i\rho(z^2\frac{d}{dz}-\tilde Nz)\cr}\right),
\ee
where we posed $N = \frac{M-1}{2}$, $\tilde N = \frac{\tilde M-1}{2}$.
The form of $\tilde H_0$ and $\tilde H_{-1}$ 
can be obtained easily.
Note that $(\tilde H_1 )_{12}=Cz$ and $(\tilde H_1 )_{21}=Dz$, 
so that $\delta=\delta' =1$ in this case.
Refering to the above general case and with
\be
\psi = \left(\matrix{\alpha_0z^n\cr
                 \beta_0z^{n}\cr}\right) + \left(\matrix{\alpha_1z^{n-1}\cr
                 \beta_1z^{n-1}\cr}\right) + \dots,
\ee
we can write the vector $\tilde H\psi$ according to
\begin{eqnarray}
&\tilde H \psi&={\rm diag}(z^{n+1},z^{n+1})M_1\left(\matrix{\alpha_0\cr\beta_0\cr}\right)\nonumber\\
&                &+\Biggl({\rm diag}(z^{n},z^{n})\tilde {M_1}\left(\matrix{\alpha_1\cr
                 \beta_1\cr}\right) + {\rm diag}(z^{n},z^{n})M_0\left(\matrix{\alpha_0\cr
                 \beta_0\cr}\right)\Biggr)\nonumber\\
&                &+ \dots,
\end{eqnarray}
where
\begin{eqnarray}
&M_1&= \left(\matrix{-4i\rho(n-N)& C\cr
                D&-4i\rho(n-\tilde N)\cr}\right),\nonumber\\
&\tilde M_1&= \left(\matrix{-4i\rho(n-1-N)& C\cr
                D&-4i\rho(n-1-\tilde N)\cr}\right),\nonumber\\
&M_0&= -(4n^2+8i\rho n+\rho^2)\unop + \left(\matrix{4i\rho N+(2N+1)^2&\tilde C\cr
        \tilde D&4i\rho \tilde N+(2\tilde N+1)^2\cr}\right).
\end{eqnarray}

The three necessary conditions for the operator $\tilde H$ to have
a finite dimensional invariant vector space can then be obtained
in a straightforward way, the final results read~:
\be
    N + \tilde N = 2 n -1 \ \ , \ \ 16 \rho^2(n-N)(n-\tilde N)+ CD = 0 \ \ ,
    \ \ \frac{\beta_0}{\alpha_0} = \frac{4 i \rho (n-N)}{C}
\ee
the equation involving the metric $M_0$ imposes in turn
\be
    \tilde C \beta_0^2 + 4 (N-\tilde N) (2 n + i \rho) \beta_0 \alpha_0 
    - \tilde D \alpha_0^2 = 0
\ee
As a result, assuming a choice of the integer $n$, we end up
with a family of QES operators labelled by the parameters
$N$, $\rho$, $C/D$ and $\tilde C$.

Different choices of the non-diagonal interactions
$H_{12}$ and $H_{21}$ can be performed which lead to
similar conditions between the cosmological constants.
We will discuss these possibilities in the framework of
periodic potentials (formulated in terms of trigonometric functions).
largely discussed in the next section.

\section{PT-symmetric QES equation with polynomial potential}
 In this section, refering to unidimensional operator studied in
 \cite{zn} 
 we will
 construct a PT-symmetric QES matrix Hamiltonian
 of the form
\be
H = -\frac{d^2}{dx^2}{\unop}_2 + M_4(x)
\ee
where $M_4(x)$ is $2\times 2$ PT-symmetric matrix.
The above Hamiltonian can be written in terms of components
and we choose the potentials of the form~:
\begin{eqnarray}
&H_{11}&= -\frac{d^2}{dx^2}-x^4+iAx^3+Bx^2+iCx+D,\nonumber\\
&H_{12}&=\omega,\nonumber\\
&H_{21}&=\tilde\omega,\nonumber\\
&H_{22}&= -\frac{d^2}{dx^2}-x^4+i\tilde Ax^3+\tilde Bx^2+i\tilde Cx+\tilde D
\end{eqnarray}
In order to reveal the QES property, it is convenient to 
perform a gauge transformation according to
\be
\tilde H = \exp(-\alpha x^3-\beta x^2 - \gamma x) H 
\exp(\alpha x^3+\beta x^2 + \gamma x),
\ee
The gauged Hamiltonian then simplifies considerably
if 
\be 
\alpha =-\frac{i}{3} \ , \ \beta=-\frac{A}{4} \ , \
\gamma = \frac{i}{2}(B-\frac{A^2}{4}) \ , \
A=\tilde A \ , \ B=\tilde B.
\ee
leading to the following expression~:
\begin{eqnarray}
&\tilde H &= -\frac{d^2}{dx^2}-4\beta x\frac{d}{dx}
-2\gamma\frac{d}{dx}-6\alpha[(x^2\frac{d}{dx}-mx)
+\theta x\sigma_3]\nonumber\\
&         &+ (-2\beta-\gamma^2) 
+ {\rm diag}(D,\tilde D) +\omega\sigma_+ + \tilde\omega\sigma_-,
\end{eqnarray}
where  the constants $C, \tilde C$ have been redefined according to
 $C=i(6\alpha (m-\theta) + 6\alpha + 4\beta\gamma)$,
 $\tilde C = i(6\alpha (m+ \theta) + 6\alpha + 4\beta\gamma)$.

However, in this form, the occurence of an invariant finite 
dimensional vector space of function is not yet manifest in the sense that
the operator $\tilde H $ doesn't preserve the vector space
$(P_{m-\theta},P_{m+\theta})^t$. In order to reveal such a possibility
we can apply the technique of the first section. Here we will follow
\cite{bh} and perform a supplementary
transformation on the operator $\tilde H$ with the matrix
$S = \left(\matrix{1&\lambda\frac{\partial}{\partial x}\cr
0&1}\right)$. After an algebra, we obtain finally the form
\begin{eqnarray}
&\hat H &= S^{-1} \tilde H S,\nonumber\\
&       &= [-\frac{d^2}{dx^2} + Ax\frac{d}{dx}-i(B-\frac{A^2}{4})\frac{d}{dx}
+ D +\frac{1}{4}(B-\frac{A^2}{4})^2]  -\tilde \omega \lambda\frac{d}{dx}\sigma_3 \nonumber\\
&       & + 2i {\rm diag}(J_+(n-2),J_+(n))
+ {\rm diag}(\frac{A}{2},-\frac{A}{2})+\tilde\omega\sigma_-
-\tilde\omega\lambda^2\frac{d^2}{dx^2}\sigma_+
\end{eqnarray}
with  $J_+(n) \equiv x^2\frac{d}{dx}-nx$. Here we have set
 $m=n-1,  \tilde D = -A+D, \theta = 1$
 and fixed the arbitrary parameter $\lambda$ entering in the gauge
 transformation   by means of $\omega = -2i\lambda n$.

The Hamiltonian $\hat H$ manifestly preserves the finite dimensional
space $(P_{n-2},P_n)^t$.
 Note that $A,B, D, \tilde\omega $ are free real parameters, $n$
is a non-negative integer and $\lambda$ is a free complex parameter.

\section{Recurence relations}
In this section we will express the formulation  of the QES solution 
in terms of recurence
relations to the case of PT-symmetic matrix Hamiltonian. 
We will see that the eigenvalue equation
$H\psi = E\psi$ leads to a system of four terms recurence relations. 
The solutions $\psi$ are of the
 form
\be
\label{ser}
\psi(x) = 
\exp(-\frac{ix^3}{3}-\frac{Ax^2}{4}+\frac{i}{2}(B-\frac{A^2}{4})x)
\left(\matrix{\sum_{k=0}^\infty P_k(E)x^k\cr
\sum_{l=0}^\infty Q_l(E)x^l\cr}\right) 
\ee
To solve the equation $H\psi = E\psi$ is equivalent to solve the 
following equation
\be
\hat H \left(\matrix{\sum_{k=0}^\infty P_k(E)x^k\cr
\sum_{l=0}^\infty Q_l(E)x^l\cr}\right) = 
E\left(\matrix{\sum_{k=0}^\infty 
P_k(E)x^k\cr\sum_{l=0}^\infty Q_l(E)x^l\cr}\right)
\ee
Then the equation above can be transformed into a {\it fourth}-order
recurence relation. It reads

\be
A_k\left(\matrix{P_k\cr
                 Q_{k+2}\cr}\right) +
                 B_k\left(\matrix{P_{k-1}\cr
                 Q_{k+1}\cr}\right)
                 + C_k\left(\matrix{P_{k-2}\cr
                 Q_k\cr}\right)
                 + D_k\left(\matrix{P_{k-3}\cr
                 Q_{k-1}\cr}\right)  = 0
\ee
where
\begin{eqnarray}
&A_k & = \left(\matrix{k(k-1)&0\cr
              -\tilde\omega & (k+2)(k+1)\cr}\right),\nonumber \\
&B_k &= \left(\matrix{[i(B-\frac{A^2}{4})+\lambda \tilde \omega](k-1)&0\cr
              0&[-\lambda \tilde \omega + i(B-\frac{A^2}{4})](k+1)\cr}\right), 
               \nonumber \\
&C_k&=\left(\matrix{-D-\frac{1}{4}(B-\frac{A^2}{4})^2
-A(k-2)-\frac{A}{2}+E&\tilde\omega\lambda^2k(k-1)\cr
   0&-D-\frac{1}{4}(B-\frac{A^2}{4})^2-Ak+\frac{A}{2}+E\cr}\right) \nonumber \\
&D_k & = -2 i \left(\matrix{(k-n-1)&0\cr
              0 & (k-n-1)\cr}\right),
\end{eqnarray}
In the present case, the recurence relations are of fourth order,
contrasting with other cases studied in the litterature \cite{bd,brihaye}
 where they are of third order. Setting $\omega=\tilde \omega = 0$
 the two recurence relations decouple and the corresponding equations
 (e.g. the one for $P_k$) correspond to the scalar PT-invariant
 and QES quartic oscillator. It is also of fourth order; as a consequence,
 both $P_0$ and $P_1$ are arbitrary ($P_0$ fixes the normalisation)
 and the other $P_k, k \geq 2$ are determined recursively.
 The construction of the QES eigenvalues associated to this system
 is not as transparant in in the case of third-order recurence
 where a common factor, say $P_n$ factorize out of the $P_k$'s, $k > n$.
 In the present case, the QES eigenvalues are obtained by solving
 the system
 \be
         P_n(E,P_1) = 0 \ \ \ , \ \ \  P_{n-1}(E,P_1) = 0
 \ee
 which is linear in $P_1$. These conditions
 indeed lead to a  truncation  of the series for $\psi_1(x)$
 defined in (\ref{ser}).
 Coming back to the full system (i.e. with
 $\omega \neq 0, \tilde \omega \neq 0$), it is easy to see that
 $Q_0, Q_1, Q_2, Q_3$ remain arbitrary ($Q_0$ set the normalisation).
 The QES eigenvalues can be obtained by solving the system
 \begin{eqnarray}
    &P_n(E,Q_1,Q_2,Q_3)= 0 \ \ , \ \   &P_{n-1}(E,Q_1,Q_2,Q_3)=0 \nonumber \\
    &Q_{n+2}(E,Q_1,Q_2,Q_3)=0 \ \ , \ \ &Q_{n+1}(E,Q_1,Q_2,Q_3)=0   \nonumber\\
 \end{eqnarray}
 which turns out to be linear in    $Q_1,Q_2,Q_3$.

\section{Conclusions}
In this paper, we have proposed a set of simple nesessary and
sufficient conditions
for matrix-valued operators of a certain type to preserve a vector space
of polynomials of fixed degrees. We have seen that the scalar
Razhavi potential admits QES matrix extensions of several types.
We also constructed a QES, matrix-valued PT invariant Hamiltonian
with  polynomial potentials. Finally, by taking this last problem
as example, we have shown that the coupled differential equations
can be transformed into a system  coupled recurence equations
 of fourth order.

\end{document}